\documentstyle[epsf,twocolumn]{article}

\newcommand{\k}{{\bf k}}
\newcommand{\q}{{\bf q}}
\begin{document}
\renewcommand{\textfraction}{0.1}
\renewcommand{\topfraction}{0.8}
\rule[-8mm]{0mm}{8mm}
\begin{minipage}[t]{16cm}
{\large \bf Effects of spin fluctuations in the $t$-$J$ model\\[4mm]}
T.~Obermeier, T.~Pruschke, J.~Keller\\[3mm]
{Institut f\"ur Theoretische Physik der Universt\"at, 93040 Regensburg, 
Germany}\\[4.5mm]
\hspace*{0.5cm}
Recent experiments on the Fermi surface and
the electronic structure
of the cuprate-supercondutors showed the
importance of short range antiferromagnetic correlations for the physics
in these systems. Theoretically, features like shadow bands
were predicted and calculated mainly for the
Hubbard model. In our approach we calculate an approximate selfenergy
of the $t$-$J$ model. Solving the $U=\infty$ Hubbard model in
the Dynamical Mean Field Theory (DMFT) yields a
selfenergy that contains most of the local correlations as a starting point.
Effects of the nearest neighbor spin interaction $J$ are
then included in a heuristical manner. Formally
like in $J$-perturbation theory all ring diagrams,
with the single bubble assumed to be purely local, are summed
to get a correction to the DMFT-self engergy
This procedure causes new bands and can furnish
strong deformation of quasiparticle bands.
\end{minipage}\\[4.5mm]
\normalsize
The obvious importance of antiferromagnetic (AF) correlations for
the physics of cuprate superconductors found an
impressing confirmation through ARUPS-experiments performed
by Aebi {\it et al.} \cite{Aeb94}. There, in a Fermi surface (FS) mapping
of $\rm Bi_2Sr_2CaCu_2O_8 (001)$ shadows of the FS were observed.
The interpretation as an effect of short range antiferromagnetic
correlations led to an intensive discussion of so called shadow features,
which have been predicted by Kampf and Schrieffer \cite{Kam90}
in a semi-phenomenological theory for the Hubbard model.

In the paper we present an approximate
evaluation of the self energy for the $t$-$J$ model
$ {\rm H}_{tJ}$ consisting of two parts:
The hopping term
\begin{equation}
{\rm H}_{t} = -\frac{t^*}{2\sqrt{2}} \sum_{{\rm <ij>}\sigma} \tilde{c}^{\dagger}_{\rm i \sigma}
\tilde{c}_{\rm j \sigma}
\end{equation}
acting only in the space of no double occupancy
($\tilde{c}^\dagger_{i\sigma}=(1-n_{i,-\sigma})c^\dagger_{i\sigma}$)
is equivalent to the $U=\infty$ Hubbard model.
The spin interaction is described explicitely by
\begin{equation}
{\rm H}_J = 
J \sum_{\rm <ij>}( \vec{S_{\rm{i}}} \cdot \vec{S_{\rm{j}}} - \frac{1}{4}
n_{\rm i} n_{\rm j}).
\end{equation}
\begin{figure}[t]
\unitlength1mm
\begin{picture}(70,68)
\end{picture}
\end{figure}

The main idea of our approach is to take into account both,
the spin fluctuations due to $ {\rm H}_J$ and the strong local correlations
in $ {\rm H}_{t}$. For the latter the dynamical mean field theory
(DMFT) \cite{Met89,Mue89,Jar93}, which becomes exact in the limit of
infinite spatial dimensions, proved to be quite a good tool.
It yields a strictly local selfenergy, which reflects the metal-insulator
transition for increasing $U$ as well as a Kondo-like resonance around
the Fermi energy. Even as an approximation for a two-dimensional system
the DMFT-results seem to be reliable when being compared to e.g.
Quantum Monte Carlo calculations \cite{Bul94,Pre95,Pru96}.
We perform the DMFT using the so called Non Crossing Approximation (NCA)
\cite{Kei70,Bic87} to solve the related local problem.
For details of this procedure see ref. \cite{Jar93}.

The quantities obtained with this method are first the local
selfenergy for calculating the approximate one-paricle Green's function
of the twodimensional $U=\infty$ Hubbard model
\begin{equation}
G_t(\k,z) = \frac{1}{z+\mu-\varepsilon_{\k}-\Sigma_{loc}(z)}
\end{equation}
and second the full local magnetic susceptibility $\chi_{loc}(\omega)$.
The dispersion is given by the Fourier transform of the next neighbor
hopping $\varepsilon_{\k}=-t^*/\sqrt{2}$ $  (\cos k_x a +\cos k_y a)$.

In addition to the correlation effects already contained in $G_t(\k,z)$,
spin fluctuations are now
included by a formal perturbation expansion in $H_J$,
taking $H_t$ as the unperturbed part
of the Hamiltonian and $G_t(\k,z)$ as the unperturbed GF ($J=0$).

As a first step
all ring diagrams are summed resulting in a RPA-like structure for the
effective interaction. In contrast to the standard RPA, however, the
intermediate polarization diagram is the {\em full} susceptibility
of the $U=\infty$ Hubbard model. Since the latter depends only very
weakly on $\q$ \cite{Pru96a} we further replace the latter quantity
by $\chi_{loc}(z)$, which can be easily calculated from the effective
local problem.

This procedure results in an effective interaction
\begin{equation}
 \tilde{V}(q)= 
	      \frac{3}{2}\frac{J(\q)}{1-J(\q)\chi_{loc}({\rm i}\nu_m)}
	 - \frac{1}{2}\frac{J(\q)}{1+J(\q)\chi_{loc}({\rm i}\nu_m)},
\end{equation}
from which we obtain a selfenergy contribution
\begin{equation}
 \Sigma_{fluct}(k)=-\sum_q G_t(k-q) \tilde{V}(q).
\end{equation}
Here we use the notation $k=(\k,{\rm i}\omega_n)$ and an implicite
sum over bosonic Matsubara frequencies
is contained in equations (4) and (5).
$\Sigma_{fluct}$ is calculated on the real frequency axis.
The $\q$-sum in eq. (5) is performed via Fast Fourier Transform on a 32x32
lattice and the sum over Matsubara frequencies by contour integration.

\begin{figure}[h]
\unitlength1cm
\begin{picture}(12,8)
\put(0,0){\epsfxsize=7.5cm\epsfbox{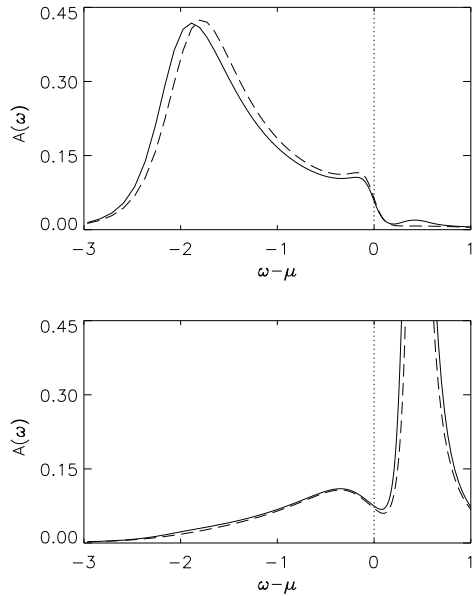}}
\end{picture}
\caption[]{
Spectral functions of the system with spin fluctuations (solid)
and without (dashed) at the $\Gamma$- and the M-point.
The solid upper curve ($\Gamma$-point) shows an additional peak, exactly
at the position of the quasiparticle peak at the M-point.
The parameters are doping $\delta=0.15$,
coupling constant $J=0.144 t^*$, inverse temperature $\beta= 10.0 t^*$.
}
\label{arpfig}
\end{figure}
The complete selfenergy is then given by
\begin{equation}
\Sigma(\k,z) = \Sigma_{loc}(z) + \Sigma_{fluct}(\k,z) + \Sigma_{Hartree},
\end{equation}
where $\Sigma_{Hartree}$ is a real number
and here determined by the occupation number of the
DMFT-result. 
This scheme is iterated with the resulting Green's function
replacing $G_t$ in
equation (5) until selfconsistency was reached.

\begin{figure}[h]
\unitlength1cm
\begin{picture}(12,8)
\put(0,-0.5){\epsfxsize=8.0cm\epsfbox{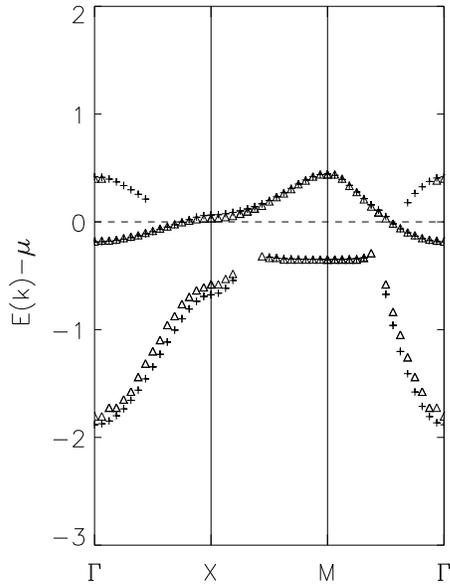}}
\end{picture}
\caption[]{
Bandstructure for the system with (crosses) and without (triangles)
spin fluctuations. Around the $\Gamma$-point a new branch is visible, which
shows the symmetry of a reduced Brillouin zone, although the system
is paramagnetic. The parameters are the same as in Fig. 1.
}
\label{bandfig}
\end{figure}

Fig. 1 shows the resulting spectral functions $A(\k,\omega)$
(full lines)
at the $\Gamma-$ and the $M-$point
(upper and lower picture), respectively
for doping $\delta=0.15$,
coupling constant $J=0.144 t^*$ and inverse temperature $\beta= 10.0 t^*$.
These two $\k$-points would be
equivalent in the case of  antiferromagnetic order.
 
In comparison with the $J=0$ case (dashed lines) an additional peak
above the Fermi energy in $A(\k_{\Gamma},\omega)$ is clearly visible,
which we interpret as a shadow of the quasiparticle peak at the M-point.
 
This becomes more clear in Fig. 2 , where the corresponding
bandstructure is plotted along high symmetry directions of the Brillouin
zone.
Crosses represent the maxima of $A(\k,\omega)$ when spin fluctuations are
taken into account. The triangles correspond to
the $J=0$ case. A new branch appears in the vicinity of the
$\Gamma$-point above the Fermi energy,
which shows the symmetry of the magnetic Brillouin zone,
as is expected for shadow bands.
As can be seen in both figures, the main features
due to strong local correlations
are not altered by the spin fluctuation effects.
 
In conclusion an extension of the dynamical mean field theory
was presented including nonlocal spin-fluctuations
in the $t$-$J$ model. With this approximation we are able to
obtain shadow features in the spectral functions.
So far our numerical calculations yield shadow bands only
above the Fermi surface. In the experiment \cite{Aeb94} shadow bands
cross the Fermi surface. We hope to reproduce this efect by
including a next-nearest neighbor hopping $t'$ in our
tight-binding bandstructure.

This work was supported by
grant number Pr 298/3-1 from the Deutsche Forschungsgemeinschaft.

\bibliography{ref}
\bibliographystyle{phys}
\end{document}